\documentstyle[12pt]{article}
\def\beq{\begin{equation}}
\def\eeq{\end{equation}}
\def\bea{\begin{eqnarray}}
\def\eea{\end{eqnarray}}
\def\bq{\begin{quote}}
\def\eq{\end{quote}}

\def\PRL{{\it Phys.Rev.Lett.} }

\def\NPB#1#2#3{{\it Nucl.\ Phys.}\/ {\bf B#1} (19#2) #3}
\def\PLB#1#2#3{{\it Phys.\ Lett.}\/ {\bf B#1} (19#2) #3}
\def\PRD#1#2#3{{\it Phys.\ Rev.}\/ {\bf D#1} (19#2) #3}
\def\PRL#1#2#3{{\it Phys.\ Rev.\ Lett.}\/ {\bf #1} (19#2) #3}

\def\MODA#1#2#3{{\it Mod.\ Phys.\ Lett.}\/ {\bf A#1} (19#2) #3}

\def\E8{$E_8\times E_8$}

\def\gappeq{\mathrel{\rlap {\raise.5ex\hbox{$>$}} {\lower.5ex\hbox{$\sim$}}}}

\def\lappeq{\mathrel{\rlap{\raise.5ex\hbox{$<$}} {\lower.5ex\hbox{$\sim$}}}}

\begin{document}
\pagestyle{empty}
\begin{flushright}
{CERN-TH/98-139}
\end{flushright}
\vspace*{5mm}
\begin{center} {\bf ASPECTS OF $M$ THEORY AND PHENOMENOLOGY}\\
\vspace*{1cm}  {\bf John ELLIS}\\
\vspace{0.3cm} Theoretical Physics Division, CERN \\ CH - 1211 Geneva 23 \\
\vspace*{2cm}   {\bf ABSTRACT} \\ 
\end{center}
A brief review is presented of selected topics, including a
world-sheet formulation of $M$ theory, couplings and scales in
$M$ phenomenology, the perils of baryon decay and the possible
elevation of free-fermion models to true $M$- or $F$-theory
compactifications.
\vspace*{5mm}
\noindent
\vspace*{1cm}

\begin{center} {\it Talk presented at the Workshop on}\\ 
{\it Phenomenological Applications of String Theories}\\ 
{\it ICTP, Trieste, October 1997} 
\end{center}
\vspace*{1.5cm}

\begin{flushleft} 
CERN-TH/98-139 \\
April 1998
\end{flushleft}
\vfill\eject

\setcounter{page}{1}
\pagestyle{plain}

\section{Perspectives on $M$ Theory}

The theories formerly known as strings have transmogrified into an  incompletely
understood higher-dimensional framework~\cite{Sen}, initially with two
formulations: $M$ theory that
energed originally in the strong-coupling limit of Type IIA string
theory, and
$F$ theory that appeared in the strong-coupling limit of Type IIB string
theory~\cite{Sen}.
There are by now several perspectives that offer clues to aspects of $M$ theory.
In the low-energy limit, it becomes 11-dimensional
supergravity. It may be
formulated on the light-cone as Matrix theory~\cite{matrix}. From a
world-sheet point of view,
$M$ theory is a dual system of world-sheet vortices and monopoles 
(representing $D$ branes in target space) close to the 
Berezinskii-Kosterlitz-Thouless
transition point~\cite{EMN}. At short distances $M$ theory may become a
topological Chern-Simons
theory with a non-compact gauge group~\cite{Horava}. 
It has recently
appeared possible to relate many of these perspectives using the
world-sheet formalism~\cite{EMNM}. According to this formulation,
string theory describes the dynamics of the Wilson loops that
characterize matter in the topological gauge theory. An enticing
aspect of this approach is the appearance of a twelfth
dimension. This may be reminiscent of $F$ theory,
which is related to $M$ theory by duality~\cite{Sen}: $M$ theory
compactified on a
space ${\cal M}$ is generally dual to $F$ theory compactified on ${\cal M}\times
S_1$. 

Among the interesting $M$ compactifications is that on $S_1/Z_2$, which
yields the strong-coupling limit of the \E8 heterotic string~\cite{W}.
This
is the perspective on $M$ theory which is most often used in phenomenological
papers, and provides the framework for most of the rest of this talk.

\section{Overview of $M$ Phenomenology}

In various different dual limits, $M$ theory manifests itself in
11-dimensional
supergravity, the \E8 heterotic string, the $SO(32)$  heterotic string, Type I,
IIA or IIB string. In the bad old days of weak-coupling string phenomenology, it
was the \E8 heterotic string that attracted the most
attention,
and the first
approaches to ${\cal M}$ phenomenology were based on the
strong-coupling version of this theory~\cite{W,HW}. It can be
formulated by considering an 11-dimensional ``bulk" space with 10-dimensional
``walls" at each of its two ends. The low-energy theory in the bulk is
11-dimensional supergravity, whilst one $E_8$ gauge theory factor appears on
each wall. This formulation appears in the strong-coupling
limit $e^{2<\phi>} \gg 1$, where $\phi$ is the dilaton in 10 dimensions. The
separation between the walls is $R_{11} \sim g^{2/3}_{string}$, which is large,
whereas the old weak-coupling limit $g_{string} \gg 1$ corresponded to
$R_{11}\rightarrow 0$ and hence a purely 10-dimensional theory.

Why did such a bizarre construction gain favour? The key phenomenological
argument is the well-known conflict between the supersymmetric grand unification
scale $m_\chi \sim 2\times 10^{16}$ GeV inferred from low-energy data obtained at
LEP and elsewhere~\cite{susyGUT}, and the string unification scale
calculated in
weakly-coupled string theory: $m_U \sim 3\times 10^{17}$
GeV~\cite{stringscale}. There were
many
pre-$M$-theory attempts to resolve this discrepancy of more than an order of
magnitude, as described below, but none of these was totally satisfactory.

The first option examined was to derive a GUT from string and break its
symmetry
at a scale $m_{GUT} \ll m_U$. Unfortunately, this is not so easy, because
the
most amenable string model constructions based on level-one realizations
of the world-sheet
current algebra could not yield the adjoint or 
other large GUT Higgs representations
required in (almost all) GUT models~\cite{noadj}. There have recently been
heroic efforts to
construct 
string GUTs using higher-level string constructions~\cite{higher},
but these are very tightly constrained, and
I am unaware of any completely convincing model.

The adjoint-Higgs problem motivated interest in flipped $SU(5)\times
U(1)$~\cite{flipped},
which only requires 10- and 5-dimensional Higgs representations that are freely
available in a low-energy effective field theory derived from a level-one string
theory. In the absence of any additional low-mass particles, the
calculated value of $\sin^2\theta_W$ may be smaller than in a conventional
$SU(5)$ GUT, if the $SU(5)$ subgroup is broken at a lower energy scale
than the string scale where it is unified with the $U(1$
factor. However,
the concordance between the measured and calculated values of
$\sin^2\theta_W$ suggests that the string unification scale at which
$SU(5)\times U(1)$ emerges cannot be much higher than the conventional GUT
scale at which $SU(5)$ is broken, so the gap remains.

This problem could in principle be resolved by a suitable coalition of light
particles, either in flipped 
$SU(5)\times U(1)$~\cite{flipinter} or some other string
model. However, the retention of the
successful prediction  for $\sin^2\theta_W$ is not automatic in such models, and
becomes a constraint rather than a glorious prediction.

Another suggestion was that the string threshold corrections might be large,
invalidating the large estimate $m_U \sim 3\times 10^{17}$ GeV. However, this
requires some moduli of the manifold of compactification to differ greatly from
the Planck scale, which is difficult to arrange in an
appealing string model.
Moreover, the successful GUT value of $\sin^2\theta_W$ again becomes a
constraint rather than a prediction~\cite{bigthresh}.

A structured approach to the possibility of additional light matter particles is
offered by the idea  that an extra space-time dimension appears at more than the
Planck length, providing many additional Kaluza-Klein states that alter the
energy-dependence of the gauge and/or gravitational couplings, and hence affect
the calculation of $m_U$. The strongly-coupled heterotic scenario for $M$
phenomenology~\cite{W} comes within this general category. In this case,
the extra
Kaluza-Klein states do not affect the running of the gauge couplings, which live
on the walls at the end of the world, but they appear in the bulk and accelerate
the running of the gravitational coupling, thereby reducing $m_U$.

It is clear that, for this scenario to work, the eleventh dimension must be
larger than $m^{-1}_{GUT}$. This makes it larger than the compactification
radius, so the sequence of events at increasing energies is $4 \rightarrow 5
\rightarrow 11$ dimensions for gravity in the bulk. Algebraically,
Newton's
constant is given by~\cite{W,Dine}
\beq
G_N = {\kappa^2\over 16\pi^2 V_6 R_{11}}
\label{one}
\eeq
where $\kappa$ is the 11-dimensional supergravity coupling: $\kappa =
(m_p^{(11)})^{9/2}$, $V_6$ is the six-dimensional compactification volume, and
the gauge coupling~\cite{W,Dine}
\beq
\alpha_{GUT} = {(4\pi \kappa^2)^{2/3}\over 2V_6}
\label{two}
\eeq
Putting in the numbers, one finds~\cite{W,Dine}
\beq
\left.
\matrix{R^{-1}_{11} \sim m_{GUT} \left({m_{GUT}\over m_P^{(11)}}\right)^2 2\pi
~\sqrt{2} ~\alpha_{GUT}^{-3/2} < m_{GUT} \cr
m_P^{(11)} = \kappa^{-2/9} = m_{GUT}~ {(4\pi)^{1/3}\over (2\alpha_{GUT})^{1/6}}
\lappeq m_{GUT}}\right\}
\label{three}
\eeq
In this scenario, there is plenty of new physics at energies below the
conventional 4-dimensional Planck scale. In particular, the spectre
appears that 5-dimensional supergravity might be the appropriate
effective field theory at energies between $R_{11}^{-1}$ and
$m_{GUT}$~\cite{MP,LOSW,ELPP}.
This could then provide the right framework for discussing the
transmission of supersymmetry breaking from the hidden wall to the
observable one~\cite{MP,ELPP}, as well as other issues~\cite{LOSW}.

The fact that $M_P$ is now a derived composite scale, and that
there is no fundamental scale much above $m_{GUT}$, provides many
phenomenological
opportunities and some challenges, one of which we now discuss, in the hope that
it may provide some inspiration for constructing interesting models derived from
$M$ theory.

\section{Caveat Baryon Decay}

The likelihood that quantum gravity might cause baryons to decay was
discussed~\cite{Hawking} before GUTS came on the scene. The basic reason
is the no-hair
theorem of quantum gravity, which indicates that the only exact symmetries are
local (gauge) symmetries. Since baryon number is only  a global quantum number
with no associated massless gauge field, one would not expect it to be
conserved. A dimension-six operator with coefficient $1/m^2_G$ would 
yield a proton lifetime $\tau_p \sim 10^{32}~(m_G/10^{15}$ GeV)$^4y$. This
would
be unobservable if $m_G \sim m_P \sim 10^{19}$ GeV: $\tau_p \sim 10^{48}
y$~\cite{Hawking}, and
would be swamped by heavy-boson exchange in conventional GUTs, since
$1/m^2_P \ll 1/m^2_{GUT}$.

In supersymmetric GUTs, there is a dimension-five mechanism for baryon decay, in
which Higgsino exchange generates an effective superpotential term of the form
$(\lambda^2/m_{\tilde H})~QQQL$~\cite{dim5}. This yields 
$(\lambda^2/m_{\tilde H}) ~(qqq\tilde q\tilde\ell , \tilde q \tilde q q
\ell )$
interactions that become
\beq
{\cal O}\left({\alpha\over
16\pi}\right)\left({\lambda^2\over m_{\tilde
H}
\tilde m}\right)~~(qqq\ell)
\eeq
interactions when dressed by sparticle loops. Thanks to the smallness of the
Yukawa couplings $\lambda$ for first-generation quarks and leptons and the loop
factors, this mechanism is at the verge of observability for $m_{\tilde H} \sim
10^{16}$ GeV~\cite{ENR,AN}.

However, what is to prevent a superpotential term $\lambda_p QQQL$ from
appearing in a
quantum theory of gravity, with $\vert \lambda_P\vert\sim 1/m_P$ ?  This
would
make baryon decay too observable: proton lifetime constraints impose
$\vert\lambda_P\vert \lappeq 10^{-6}/m_P$~\cite{grav5}.

We have studied this question in some specific models derived from
string~\cite{ELN}. For
example, no such dimension-five operators are generated by $\tilde H$ eschange
in the effective field theory. Moreover, the coefficients of non-renormalizable
superpotential terms are calculable in any given model, and may be absent in
some specific models, as a result of $U(1)$ or other symmetries.

The problem becomes more acute in the new $M$-phenomenology
framework~\cite{Dine,EFN}. Consider
a generic non-renormalizable interaction
\beq
0(1) \times {g^{N+2}_{string}\over M^{N+1}} ~~QQQL \Phi^N
\label{four}
\eeq
The natural scale in the denominator is now $M\rightarrow m_{GUT} \sim 10^{16}$
GeV rather than $m_P\sim 10^{19}$ GeV, one can expect $<\Phi > /M\sim
0(1)$ in
general, and the string coupling in the numerator is $0(1)$. We need some
powerful symmetry or other principle to suppress such operators, perhaps to all
orders in perturbation theory and at the non-perturbative level.

We have approached this problem~\cite{EFN} using one of the available
technologies for
string model-building, derived in the context of weakly-coupled heterotic
string.
In this way we may identify models with a chance of suppressing baryon decay,
that one may be elevate to the strong-coupling limit of $M(F)$ theory.

\section{Building Models with Free Fermions}

This approach starts from free fermions on the world sheet~\cite{freef},
which are divided
into sets $b_k$ with specified  boundary conditions $f\rightarrow
e^{i\alpha_k}f$, forming a finite additive group $\Xi$. The physical states in a
given sector $\xi \in \Xi$ are then obtained by making generalized GSO
projections. The choices of boundary conditions are subject to many consistency
conditions imposed by modular invariance. There are $U(1)$ charges $Q(f) =
{1\over 2} \alpha(f) + F(f)$, where $\alpha(f)$ is the fermionic boundary
condition for $F$, and $F(f) = \pm 1$ for $f, f^\star$. These may be enhanced to
non-Abelian symmetries by appropriate choices of the boundary conditions. 

Many interesting models are derived by starting with a particular set (called
NAHE) of five boundary-condition vectors $\{\mbox{\boldmath$1$}, S, b_1, b_2,
b_3\}$, which yield after the generalized GSO projections an $N = 1$
supersymmetric $SO(1)\times SO(6)^3\times E_8$ gauge group~\cite{NAHE}.
The
boundary-condition vectors $\{b_1,b_2,b_3\}$ provide three twisted sectors that
each yield 16  $\underline{16}$  representations of $SO(10)$. The models are
differentiated by their choices of additional basis vectors, which reduce the
spectrum to three generations with an observable-sector gauge group that may be
$SU(5)\times U(1)$~\cite{flipped}, $SO(6)\times SO(4)$~\cite{sixfour} or
$SU(3)\times
SU(2)\times U(1)^2$~\cite{Faraggi},
with extra observable-sector Higgs representations in a
$\underline{16}+\underline{16}$ of $SO(10)$, and a hidden sector gauge group
that is a subgroup of $E_8$, and has matter representations in general.

We have studied~\cite{EFN} two specific models to see whether they avoid
the
baryon-stability problem. The first of these has~\cite{model1} two
dangerous sixth-order terms
in the superpotential of the forms ${1\over M^3} QQQL~\Phi \bar\Phi$, that would
yield dangerous dimension-five proton decay operators if 
${<\Phi>~<\bar\Phi>\not= 0}$. Some such vacuum expectation values are
necessarily generated by an anomalous $U(1)$, and we can expect the flatness
conditions on the potential to generate 
\beq
{<\Phi>~<\bar\Phi>\not= 0}
\eeq
In the
$M$-theory context, we do not expect these vacuum expectation values to be
much smaller than the
mass scale $M\sim 10^{16}$ GeV. Moreover, many analogous operators appear at
higher orders. Therefore, this model exemplifies the generic problems we expect
with baryon stability in $M$ theory.

A more promising model~\cite{model2} is one with an enhanced gauge
symmetry. It has a
Neveu-Schwarz sector that yields an $SU(3)\times SU(2)\times U(1)_C\times
U(1)_L\times U(1)^6$ gauge group. Suitable further choices of boundary
conditions elevate one particular combination of $U(1)_C, U(1)_L, \cdots$ to
become an $SU(2)$ gauge group. The conventional electric charge $Q_{em} = T^3_L
+ Y = T^3_L + \hat Y + {1\over 2} T^3_{cust}$, with a component in this
custodial $SU(2)$ symmetry group. Among the conventional three light
generations, the leptons $L$ and $e^C_L$ are $SU(2)_{cust}$ doublets, whilst the
quarks $Q, u^C_L, d^C_L$ are $SU(2)_{cust}$ singlets. This immediately implies
that there are no $QQQL$ terms, and the quantum numbers of the candidate
$\Phi, \bar\Phi$ fields ensure that no such terms are generated in any order of
perturbation theory~\cite{EFN}.

This is certainly a promising start, though there is no guaranted what
non-perturbative effects may appear in $M$ theory. On the other hand, a generic
$M$-theory model may even possess additional non-perturbative gauge symmetries.
If one stays within the weak-coupling free-fermion approach, the strategy is to
add to the NAHE set of basis vectors some new vector $\gamma$ which, in
combination with others, yields a sector containing additional massless
space-time gauge bosons. However, the quarks and leptons must transform
non-trivially under this enhanced symmetry. It is no good if it only acts on
hidden-sector states, for example, and this depends on details of the GSO
projection.

This motivates a search for more powerful analysis tools that may be elevated to
the full $M(F)$-theory context. At the moment, we are unaware of a general
strong-coupling equivalent of the free-fermion model. On the other hand, some
geometric tools for compactifying $M(F)$ theory have been developed. Hence it is
desirable, as a first step, to understand the geometry underlying the NAHE set
of boundary conditions~\cite{BEFNQ}.
We have first identified how the NAHE set may be found within the general
class of $Z_2\times Z_2$ orbifolds. The NAHE free-fermion point
corresponds to a compactified lattice with enhanced $SO(12)$
symmetry, rather than the more familiar $Z_2 \times Z_2$ orbifold
that is based on a $(T_2)^3$ Narain lattice with $SO(4)^3$ symmetry
The NAHE model has Euler
characteristic $\chi = 48$, with $h_{11} = 27, h_{21} = 3$, 
whereas the more familiar model has $h_{11} = 51, h_{21} = 3$.

\section{Connection to $M(F)$ Compactifications?}

There is rather little literature on these. Much of it concerns $Z_2\times Z_2$
orientifolds related to the orbifold model with $h_{11} = 51, h_{21} =
3$~\cite{MV}.
Known $M(F)$ theory
compactifications on Calabi-Yau threefolds may be classified in terms of three
invariants $(r, a, \delta) : h_{11} = 5+3r -2a$, $h_{21} = 65-3r-2a$. There is a
limited catalogue of such models due to Voisin, Borcea and
Nikulin~\cite{VBN}. The
``standard" $Z_2\times Z_2$ orbifold model has $(r,a,\delta ) = (18,4,0)$, whilst
the NAHE orbifold ``should" have $(r,a,\delta ) = (14,10,0)$, which is not in the
catalogue!

We are currently looking directly for a Calabi-Yau threefold
compactification that corresponds to
the NAHE set. A convenient way to tackle this problem is to use the
Landau-Ginzburg formalism. We have found~\cite{BEFNQ} an interesting class
of such models based on
the superpotential
\beq
W = X^4_1 + X^4_2 + X^2_3 + X^4_4 + X^4_5 + X^2_6 + X^4_7 + X^4_8 + X^2_9
\eeq
with
spectra modded out by discrete symmetries. We have identified one such model
that has $(h_{11},h_{21}) = (51,3)$ and 
has the right symmetries to correspond to the known Voisin-Borcea
model~\cite{VBN,MV}, and
another that has $(h_{11},h_{21}) = (27,3)$ and
apparently corresponds to the NAHE set.~\footnote{Other members of this
family of models include the morrors of these examples.} 
If so, the next step will be to see whether it yields a consistent extension of
the known Voisin-Borcea models, and in particular whether it an
elliptic fibration as sought for $M$- and $F$-theory
compactifications~\cite{MV}. If so, we would
have a consistent $M(F)$ elevation of
the NAHE free-fermion models, that may provide new phenomenological insights
into issues such as proton decay.

\section{Conclusions}

$M$ phenomenology is very much in its infancy. Although impressive technical
progress in $M$ and $F$ theory has been achieved, little progress has yet been
made on the construction of interesting models. Some purists would consider any
such effort premature before all the theoretical problems have been resolved.
Perhaps theoretical consistency will even determine uniquely the choice of
vacuum. I disagree. I believe that experiment surely has a r\^ole to play, and
think that it is useful to pursue a complementary bottom-up approach that uses
our empirical knowledge in an attempt to figure out aspects of the Big Picture
even before all the pieces of the theoretical jigsaw are in place.

Baryon stability may be one of the important clues. It was already a headache
for compactifications of the weakly-coupled heterotic string. It may be a more
serious problem in $M$ theory, with its agglomeration of physics scales around
$m_{GUT} \simeq 10^{16}$ GeV $ \ll m_P$~\cite{EFN}.  Free-fermion fmodels
may provide
a
useful tool for analyzing this problem, and models based on the NAHE structure
with an enhanced gauge  symmetry may be particularly promising. Elevating
these
models to true $M(F)$ theory compactifications requires more geometric
intuition. The NAHE set would correspond to some generalization of the known
Voisin-Borcea models, and some progress towards identifying this seems
to be emerging~\cite{BEFNQ}.

Looking beyond this horizon, non-perturbative string theory offers many
further prospects for model building that had not previously been
considered~\cite{Ibanez}. These include possible enlargements of the gauge
group to
a rank (considerably) larger than 22 and the treatment of transitions that
change the number of chiral fields, as well as duality relations
between strong- and weakly-coupled models. It will take time to
learn how to apply all these tricks to the construction of
realistic phenomenological models, but they offer exciting
prospects, such as the hope of determining dynamically the
number of generations. It will be interesting to see whether the
ultimate string- or $M$-theory model bears a close relation
to the specific models that have been studied up to now. Very likely not,
but I believe that at least some of the 
phenomenological lessons we have learnt 
may stand us in good stead as we search for this ultimate model.

\end{document}